\documentstyle[12pt,grapcxpc]{article}
\begin{document}

\begin{titlepage}
\normalsize
\begin{center}
{\Large \bf Budker Institute of Nuclear Physics}
\end{center}
\begin{flushright}
BINP 94-56\\
June 1994
\end{flushright}
\vspace{1.0cm}
\begin{center}
{\bf NUCLEAR-SPIN-DEPENDENT PARITY-NONCONSERVING EFFECTS}
\end{center}

\begin{center}
{\bf IN THALLIUM, LEAD AND BISMUTH ATOMS}
\end{center}

\vspace{1.0cm}
\begin{center}
{\bf I.B. Khriplovich}\footnote{e-mail address: khriplovich@inp.nsk.su}
\end{center}

\begin{center}
Budker Institute of Nuclear Physics, 630090 Novosibirsk,
Russia
\vspace{4.0cm}
\end{center}

\begin{abstract}
Nuclear-spin-dependent P-odd optical activity in atomic Tl, Pb and Bi is
calculated. Its magnitude is expressed analytically through the main
contribution to the optical rotation, which is independent of nuclear spin.
The accuracy of results is discussed.
\end{abstract}

\end{titlepage}

\section{Introduction}
Up to now only parity-nonconserving (PNC) effects independent of nuclear spin
have been reliably observed in atomic experiments. The reason is that in heavy
atoms, investigated experimentally, P-odd nuclear-spin-dependent (NSD)
correlations are smaller than those observed, roughly by two orders of
magnitude.

In heavy atoms the NSD P-odd effects were shown \cite{fk,fks} to be induced
mainly by the anapole moment of a nucleus, its P-odd electromagnetic
characteristic. The measurements of these effects would give valuable
information on PNC nuclear forces.

P-odd interaction with nuclear spin leads to some difference in the
magnitude of PNC effects at different hyperfine components of optical
transitions \cite{nsfk}. Experiments aimed at the detection of NSD P-odd
effects in cesium, thallium, lead, bismuth are underway in many groups.
The first evidence of those correlations has been seen (at the
level of two standard deviations) in cesium \cite{wie}. Meaningful upper
limits on the NSD optical activity in bismuth and lead were obtained in
Refs.\cite{mzw,mvm}. Recently the accuracy close to that necessary for the
detection of NSD effects has been achieved in the thallium optical
rotation experiment \cite{lam}.

The first atomic calculations of the NSD P-odd effects were presented in
Ref.\cite{nsfk}. More careful and accurate calculations of those
correlations in cesium were carried out in Refs.\cite{frk,kra,bjs,bp}, the
results of those last works being in good mutual agreement.

In view of the mentioned optical rotation experiments \cite{mzw,mvm,lam}, a
theoretical investigation of NSD PNC effects in thallium, lead and bismuth,
alternative to that of Ref.\cite{nsfk}, is relevant and timely. This is the
subject of the present paper.

\section{Thallium}
The simplest case is that of the $6p_{1/2}\longrightarrow 6p_{3/2}$
transition in thallium. PNC weak interaction $\hat{W}$ admixes the states
$ns_{1/2}$ to the initial one, and the mixing matrix element is
proportional to (see, e.g., Ref.\cite{khr})
\begin{equation}\label{mix}
<s_{1/2}|\hat{W}|p_{1/2}> \sim i \left\{\frac{Q}{2}
+ \kappa_a \frac{2\gamma + 1}{3} 2{\bf j}{\bf i} \frac{K}{i(i+1)}\right\}.
\end{equation}
The weak nuclear charge $Q$ in this expression is close numerically to
$-N$, $N$ being the neutron number; $N=122$ and $124$ for $^{203}$Tl and
$^{205}$Tl respectively. The dimensionless anapole constant $\kappa_a$ is
close in Tl to 0.40 \cite{fks,dkt}. As to other notations in expression
(\ref{mix}), $j=1/2$ is the electron angular momentum, $i$ is the nuclear
spin, $K=(l-i)(2i+1), \;l$ is the orbital angular momentum of the valence
nucleon (for both thallium isotopes $i=1/2,\; K=-1,\; l=0$);
$\gamma=\sqrt{1-Z^2\alpha^2}$. An overall numerical factor omitted in the
rhs of formula (\ref{mix}) is irrelevant for our treatment; the imaginary
value of matrix element (\ref{mix}) is essential for the transition to the
second-quantization representation (see below). Due to the short-range nature
of the PNC electron-nucleus interaction $\hat{W}$, mixing of any other pair of
opposite-parity one-electron states can be neglected.

There are other contributions to the NSD term in expression (\ref{mix}), due to
the weak neutral currents, and to the combined effect of the weak neutral
charge $Q$ and usual hyperfine interaction \cite{fk1,koz,bp1}. Those
contributions can be easily included into the consideration, but most probably
they are numerically small as compared to $\kappa_a$.

Mixing (\ref{mix}) conserves obviously the total atomic angular momentum $F$
which in this case refers to the initial state $6p_{1/2}$, and the matrix
element reduces to
\begin{eqnarray}\label{rat}
\frac{Q}{2} + \kappa_a \frac{2\gamma + 1}{3} \left\{ \begin{array}{cl}
                                                         2, & F=0\\
                                                       -2/3,& F=1
                                                       \end{array}
\right. .
\end{eqnarray}

The second type of excitation essential here is that of electrons belonging to
the $6s^2$ subshell: the states of the configuration $6s_{1/2} 6p_{1/2}
6p_{3/2}$ are admixed by the weak interaction to $6s^2 6p_{3/2}$. This
contribution can be formally described as a transition of the $6p_{1/2}$ into
the occupied $6s_{1/2}$ state, induced by the weak interaction, with the
subsequent E1 transition from $6s_{1/2}$ to $6p_{3/2}$. No wonder therefore
that the ratio between the nuclear-spin-independent (NSI) effect due to $Q$ and
that driven by the anapole moment (AM) interaction, proportional to $\kappa_a$,
is controlled by the same relation (\ref{rat}) where $F$ refers again to the
initial state. This result can be as well obtained directly using the
second-quantization representation and the completeness relation for the states
belonging to the configuration $6s_{1/2} 6p_{1/2} 6p_{3/2}$ (see below).

In this way we obtain the following relation between the circular polarization
$P_a$ induced by the nuclear AM and that induced by the weak nuclear
charge $Q$:
\begin{eqnarray}
\frac{P_a}{\kappa_a}=\frac{P_Q}{Q}
\frac{2\gamma + 1}{3} \left\{ \begin{array}{cl}
                                      4, & F=0\\
                                    -4/3,& F=1
                              \end{array}
                                 \right. .
\end{eqnarray}

Taking for $P_Q$ the result obtained in the phenomenological approach
\cite{nsfk1,khr}
\begin{equation}\label{phth}
P_Q = 3.4\cdot 10^{-7}(-Q/N),
\end{equation}
we get finally
\begin{eqnarray}\label{th}
\frac{P_a}{\kappa_a} = \left\{ \begin{array}{rl}
  -0.96\cdot 10^{-8},  & F=0\longrightarrow F^{\prime}=1\\
   0.32\cdot 10^{-8},  & F=1\longrightarrow F^{\prime}=1,\;2
   \end{array}
   \right. .
\end{eqnarray}

Not only the result of Ref.\cite{nsfk} is reproduced\footnote{In paper
\cite{nsfk} we used different normalization for the NSD effect, natural for
the neutral current interaction. For comparison with the present results (and
with the numbers quoted in book \cite{khr}) the original ones should be
multiplied by $(1-2K)/2K$. In thallium this factor equals $-3/2$.}, but it is
clear now why for both transitions from the state of $F=1$ the degree of
circular polarization is the same.

In conclusion of this section it should be mentioned that relation
(\ref{mix}) allows one to get immediately the magnitude of the NSD PNC
effect in the strongly forbidden transition $6p_{1/2}\longrightarrow
7p_{1/2}$ in thallium at $F^{\prime}=F=1$, again in complete agreement
with the result of Ref.\cite{nsfk}. This trick was applied previously in
Ref.\cite{frk} (see also \cite{bp}) to the components with $F^{\prime}=F$
of the transition $6s_{1/2}\longrightarrow 7s_{1/2}$ in cesium.

\section{Lead}
The next case of experimental interest is the $^3 P^{\prime}_0 \longrightarrow
^{3}P_1$ transition in the odd isotope of lead, $^{207}$ Pb ($i=1/2,\;K=1$).

It is convenient here to use formula (\ref{mix}) in the second-quantization
representation:
\begin{eqnarray}\label{msq}
\hat{W} \sim i \left \{\frac{Q}{2}(a^{\dagger}_{+}b_{+}+a^{\dagger}_{-}b_{-}
- b^{\dagger}_{+}a_{+} - b^{\dagger}_{-}a_{-})\right.\nonumber \\
+ q [i_{+}(a^{\dagger}_{-}b_{+} - b^{\dagger}_{-}a_{+})
+ i_{-} (a^{\dagger}_{+}b_{-} - b^{\dagger}_{+}a_{-})\nonumber \\
+ i_z (a^{\dagger}_{+}b_{+} - a^{\dagger}_{-}b_{-} -b^{\dagger}_{+}a_{+}
+ b^{\dagger}_{-}a_{-})] \left.\right \}.
\end{eqnarray}
Here $a^{\dagger}\;(a)$ and $b^{\dagger}\;(b)$ are the creation (annihilation)
operators of the $s_{1/2}$ and $p_{1/2}$ electrons respectively, the subscripts
refer to the signs of electron angular momentum projections $j_z = \pm 1/2$;
$$q = \kappa_a \frac{2\gamma + 1}{3} \frac{K}{i(i+1)}.$$

We will need also the second-quantization representation for the operator of E1
$s-p$ transition. Up to an overall factor, which is again irrelevant for our
purpose, this operator is (see, e.g., Ref.\cite{khr}):
\begin{eqnarray}\label{e1}
\hat{D}^{+} \sim - i \left \{\rho_{1/2}\sqrt{2}(a^{\dagger}_{+}b_{-}+
b^{\dagger}_{+}a_{-})\right.\nonumber \\
+ \rho_{3/2}(\sqrt{3}c^{\dagger}_{3/2}a_{+} + c^{\dagger}_{1/2}a_{-}
 - \sqrt{3}a^{\dagger}_{-}c_{-3/2} - a^{\dagger}_{+}c_{-1/2})\left.\right \}.
\end{eqnarray}
Here $c^{\dagger}_m\;(c_m)$ is the creation (annihilation)
operator of the $p_{3/2}$ electron with $j_z=m$; $\rho_j$ is the scalar radial
integral of E1 $s-p_{j}$ transition.

The ground state wave function of the lead $6p^2$ configuration is expanded as
follows in the $jj$ basis (see, e.g., Ref.\cite{khr}):
\begin{equation}\label{intc}
|^3 P^{\prime}_0>=a_1 |1/2\;1/2>_0 + a_2 |3/2\;3/2>_0;\;\;
a_1=0.974,\;a_2=-0.227.
\end{equation}
The symbol $|j_1\;j_2>_J$ denotes the normalized wave function of two
p-electrons with angular momenta $j_1$ and $j_2$, and total angular momentum
$J$. The first excited state of that configuration is pure both in the LS and
$jj$-scheme:
\begin{equation}
|^3 P_1>= |1/2\;3/2>_1.
\end{equation}
In the second-quantization representation the basic wave functions are
\begin{equation}
|1/2\;1/2>_0 = b^{\dagger}_{+} b^{\dagger}_{-}|0>,
\end{equation}
\begin{equation}
|3/2\;3/2>_0 =\frac{1}{\sqrt{2}} (c^{\dagger}_{3/2}c^{\dagger}_{-3/2}
- c^{\dagger}_{1/2}c^{\dagger}_{-1/2})|0>,
\end{equation}
\begin{equation}
|1/2\;3/2>_1 = \frac{1}{2} (\sqrt{3} c^{\dagger}_{3/2} b^{\dagger}_{-}
- c^{\dagger}_{1/2}b^{\dagger}_{+}|0>.
\end{equation}
Here $|0>$ is the wave function of the closed shells.

The admixed E1 transition from one of the ground state $jj$ components,
$|1/2\;1/2>_0$, to $|1/2\;3/2>_1$ proceeds via states of the type
$ns_{1/2}6p_{1/2}$ (including the subshell $6s^2$ excitation, as it was the
case in thallium) and is described by the effective operator
$\hat{D}^{+}\hat{W}$. Small spin $i=1/2$ of the isotope $^{207}$Pb simplifies
the calculations essentially. The total atomic angular momentum of the ground
state is fixed, $F=1/2$, and coincides with the nuclear spin. Elementary
calculations demonstrate that
\begin{equation}
\hat{D}^{+}\hat{W} |1/2\;1/2>_0 |->
\sim (Q - q) \sqrt{2/3}\, |1/2,1/2)
+ (Q + q/2) \sqrt{1/3}\, |3/2,1/2).
\end{equation}
Here $|->$ denotes the nuclear spin state with $i_z=-1/2$ (at $J=0$ it
corresponds to $F_z=-1/2$), $\;|F,F_z)$ refers to atomic eigenstate of given
$F$ and $F_z$.

The transition from the second ground state component,
$|3/2\;3/2>_0 \longrightarrow |1/2\;3/2>_1$, corresponds to the effective
operator $\hat{W} \hat{D}^{+}$. As easily one gets
\begin{equation}
\hat{W} \hat{D}^{+} |3/2\;3/2>_0 |->
\sim (Q + q) \sqrt{1/3}\, |1/2,1/2)
+ (Q - q/2) \sqrt{2/3}\, |3/2,1/2).
\end{equation}

Therefore, the admixed E1 amplitude of the $F=1/2\longrightarrow
F^{\prime}=1/2$ transition is proportional to
$$Q(a_1\sqrt{2/3}\,+a_2\sqrt{1/3}\,)
 - q(a_1\sqrt{2/3}\,-a_2\sqrt{1/3}\,);$$
that of $F=1/2\longrightarrow F^{\prime}=3/2$ to
$$Q(a_1\sqrt{1/3}\,+a_2\sqrt{1/6}\,)
 + q/2(a_1\sqrt{1/3}\,-a_2\sqrt{1/6}\,).$$
Finally, we get
\begin{eqnarray}
\frac{P_a}{\kappa_a}=\frac{P_Q}{Q}\;
\frac{2\gamma + 1}{3}\;\frac{2}{3}\;
\frac{1 - a_2/(a_1\sqrt{2})}{1 + a_2/(a_1\sqrt{2})}
                   \left\{ \begin{array}{cl}
                                      -2, & F^{\prime}=1/2\\
                                       1, & F^{\prime}=3/2
                              \end{array}
                                 \right. .
\end{eqnarray}
Taking again for $P_Q$ the result of the phenomenological calculations
\cite{nsfk1,khr}
\begin{equation}\label{phle}
P_Q = 2.4\cdot 10^{-7}(-Q/N),
\end{equation}
we come to the following numerical predictions:
\begin{eqnarray}\label{le}
\frac{P_a}{\kappa_a} = \left\{ \begin{array}{rl}
   0.31\cdot 10^{-8},  & F^{\prime}=1/2\\
  -0.15\cdot 10^{-8},  & F^{\prime}=3/2
   \end{array}
   \right.
\end{eqnarray}
in complete agreement with Ref.\cite{nsfk}.

\section{Bismuth}
Of experimental interest in bismuth are the transitions from the ground state
into the first and second excited ones. All those levels belong to the
configuration $6p^3$. The ground state and the first excited one have $J=3/2$.
The coefficients of their expansion in $jj$ basis
$$a_1|1/2\;1/2\;3/2>_{3/2}+a_2|1/2\;3/2\;3/2>_{3/2}
+a_3|3/2\;3/2\;3/2>_{3/2},$$
as derived in Ref.\cite{sfk1} (and quoted in Ref.\cite{khr}), are presented in
Table 1. Their standard values \cite{ll} are given in brackets in the same
Table. The second excited state is pure $|1/2/;3/2/;3/2>_{5/2}$.

  \begin{table}[h]
  \begin{center}
  \begin{tabular}{ccccc}
                    &   &     $a_1$       &    $a_2$        &     $a_3$      \\
                    &   &                 &                 &                \\
Ground state        &&$-$0.929 ($-$0.935) & 0.323 (0.308)&$-$0.179 ($-$0.172)\\
                    &   &                 &                 &              \\
First excited state &&$-$0.336 ($-$0.324)&$-$0.940 ($-$0.944) & 0.053 (0.066)\\
        \end{tabular}
  \end{center}
  \caption{Expansion of bismuth $6p^3$ states in $jj$ basis}
  \label{ic}
  \end{table}

In the second-quantization representation the basic $jj$ functions with maximum
projections are:
\begin{equation}\label{1}
|1/2\;1/2\;3/2>_{3/2}=c^{\dagger}_{3/2}b^{\dagger}_{+}b^{\dagger}_{-}|0>,
\end{equation}
\begin{equation}\label{2}
|1/2\;3/2\;3/2>_{3/2}=\frac{1}{\sqrt{5}}c^{\dagger}_{3/2}(2c^{\dagger}_{1/2}
b^{\dagger}_{-} - c^{\dagger}_{-1/2}b^{\dagger}_{+})|0>,
\end{equation}
\begin{equation}\label{3}
|3/2\;3/2\;3/2>_{3/2}=c^{\dagger}_{3/2}c^{\dagger}_{1/2}c^{\dagger}_{-1/2}|0>,
\end{equation}
\begin{equation}\label{4}
|1/2\;3/2\;3/2>_{5/2}=c^{\dagger}_{3/2}c^{\dagger}_{1/2}b^{\dagger}_{+})|0>.
\end{equation}

The only stable bismuth isotope $^{209}$Bi has spin $i=9/2$ ($K=5$). Due to
large $J$'s and $i$ both transitions from the ground state, into the first
excited one (876 nm) and into the second (648 nm), have a lot of hyperfine
components, 10 and 12 respectively. This makes us to resort, instead of the
elementary treatment of lead and especially of thallium, to heavy artillery of
$3nj$ symbols.

In the single-electron language the following chains contribute to both
transitions:
$$p_{1/2}\longrightarrow s_{1/2}\longrightarrow p_{1/2},$$
$$p_{1/2}\longrightarrow s_{1/2}\longrightarrow p_{3/2},$$
$$p_{3/2}\longrightarrow s_{1/2}\longrightarrow p_{1/2}.$$
The first of them is described by the effective operator
$\hat{D}^{+}\hat{W}+\hat{W}\hat{D}^{+}$ which reduces in this case to
\begin{equation}\label{ch1}
\hat{D}^{+}\hat{W}+\hat{W}\hat{D}^{+} \sim
q\sqrt{2}\{i_{+}(b^{\dagger}_{+}b_{+} - b^{\dagger}_{-}b_{-})
-2i_z\,b^{\dagger}_{+}b_{-}\} = q\,2\sqrt{2}[\hat{V}_1\times i]^1_{+}.
\end{equation}
In this expression vector
\begin{equation}
\hat{V}^1= \{-\sqrt{2}b^{\dagger}_{+}b_{-},
b^{\dagger}_{+}b_{+} - b^{\dagger}_{-}b_{-}, \sqrt{2}b^{\dagger}_{-}b_{+}\}
\end{equation}
is nothing else but the operator 2{\bf j} for $p_{1/2}$ electron in the
second-quantization representation; the symbol $[\hat{V}^1\times i]^1_{+}$
denotes as usual the tensor product of two operators coupled into a vector with
projection $+1$.

The effective operator for the second chain is
\begin{eqnarray}
\hat{D}^{+}\hat{W} \sim
Q \frac{1}{2} (\sqrt{3}\,c^{\dagger}_{3/2}b_{+}
+ c^{\dagger}_{1/2}b_{-})\nonumber \\
+ q [i_{+} c^{\dagger}_{1/2}b_{+} + i_{-} c^{\dagger}_{3/2}b_{-}\sqrt{3}
+ i_z (\sqrt{3}c^{\dagger}_{3/2}b_{+} - c^{\dagger}_{1/2}b_{-})]\nonumber \\
= Q \,\hat{V}_{+} + q\,\{2\sqrt{2}[\hat{V}\times i]^1_{+}
+ \sqrt{10} [\hat{T}\times i]^1_{+}\}.
\end{eqnarray}
The tensor operator $\hat{T}$ is defined in such a way that
$$\hat{T}_{+2} = c^{\dagger}_{3/2}b_{-}.$$
The definition of the vector operator $\hat{V}$ is clear from the equation
itself.

Quite analogously the third chain is described by the effective operator
\begin{eqnarray}
\hat{W}\hat{D}^{+}\longrightarrow
Q \frac{1}{2} (\sqrt{3}\,b{\dagger}_{-}c_{-3/2}
+ b{\dagger}_{+}c_{-1/2})\nonumber \\
+ q [i_{+} b{\dagger}_{+}c_{1/2} + i_{-} b{\dagger}_{-}c_{3/2}\sqrt{3}
+ i_z (-\sqrt{3}b{\dagger}_{-}c_{-3/2} + b{\dagger}_{+}c_{-1/2})]\nonumber \\
= Q \,\hat{\tilde{V}}_{+} + q\,\{-2\sqrt{2}[\hat{\tilde{V}}\times i]^1_{+}
+ \sqrt{10} [\hat{\tilde{T}}\times i]^1_{+}\}
\end{eqnarray}
where
$$\hat{\tilde{T}}_{+2} = b^{\dagger}_{+}c_{-3/2}.$$

Now the standard technique of the angular momentum theory allows one to
obtain closed expressions for the reduced matrix elements of those operators
between the hyperfine states. For the infrared transition it is
\begin{eqnarray}\label{ik}
\sqrt{(2F^{\prime}+1)(2F+1)}\Biggl[(-)^{F+1}
                   \left\{ \begin{array}{ccc}
                                      3/2        & F^{\prime} & i \\
                                      F          & 3/2        & 1
                              \end{array}
                              \right \}
Q\sqrt{6} (a_1 a^{\prime}_2 +a_2 a^{\prime}_3 - a_2 a^{\prime}_1
- a_3 a^{\prime}_2)\nonumber \\
+ q \sqrt{i(i+1)(2i+1)}\Biggl(\left\{ \begin{array}{ccc}
                                      3/2        &  3/2 & 1 \\
                                      i          &  i   & 1 \\
                                      F^{\prime} &  F   & 1
                              \end{array}
                              \right \}
6(a_1 a^{\prime}_2 +a_2 a^{\prime}_3 + a_2 a^{\prime}_1 + a_3 a^{\prime}_2
- 2\sqrt{2/5}\, a_2 a^{\prime}_2)\nonumber \\
+ \left\{ \begin{array}{ccc}
                                      3/2        &  3/2 & 2 \\
                                      i          &  i   & 1 \\
                                      F^{\prime} &  F   & 1
                              \end{array}
                              \right \}
(-2\sqrt{15}) (a_1 a^{\prime}_2 - a_2 a^{\prime}_3 - a_2 a^{\prime}_1
+ a_3 a^{\prime}_2)\Biggr)\Biggr]
\end{eqnarray}
Here unprimed and primed factors $a_i$ refer to the ground and first excited
states respectively.  For the red transition the analogous formula is
\begin{eqnarray}\label{k}
 - \sqrt{(2F^{\prime}+1)(2F+1)}\Biggl[(-)^{F}
                   \left\{ \begin{array}{ccc}
                                      5/2        & F^{\prime} & i \\
                                      F          & 3/2        & 1
                              \end{array}
                              \right \}
Q\sqrt{3/2}\, (a_1 + a_3)\nonumber \\
+ 3q \sqrt{i(i+1)(2i+1)}\Biggl(\left\{ \begin{array}{ccc}
                                      5/2        &  3/2 & 1 \\
                                      i          &  i   & 1 \\
                                      F^{\prime} &  F   & 1
                              \end{array}
                              \right \}
(a_1 + a_3 + 8\sqrt{2/5}\, a_2)\nonumber \\
+ \left\{ \begin{array}{ccc}
                                      5/2        &  3/2 & 2 \\
                                      i          &  i   & 1 \\
                                      F^{\prime} &  F   & 1
                              \end{array}
                              \right \}
\sqrt{35}\,(a_1 + a_3)\Biggr)\Biggr]
{}.
\end{eqnarray}
We have used in both formulae (\ref{ik}) and (\ref{k}) the explicit values of
the reduced matrix elements of the operators $V^1,\;V,\;\tilde{V},\;T$ and
$\tilde{T}$ between the $jj$ states (\ref{1}) - (\ref{4}) which can be found
easily.

We are interested in the ratios of the NSD terms in those matrix elements,
which are proportional to $q$, to NSI ones, which are proportional to $Q$. The
following identity is useful here\cite{vmkh}:
\begin{eqnarray}
              \sqrt{i(i+1)(2i+1)} \left\{\begin{array}{ccc}
                                      j^{\prime} &  j   & 1 \\
                                      i          &  i   & 1 \\
                                      F^{\prime} &  F   & 1
                              \end{array}
                              \right\}\nonumber \\
=\frac{(F^{\prime} - j^{\prime})(F^{\prime} + j^{\prime} +1)
- (F-j)(F+j+1)}{2\sqrt{6}}(-)^{j^{\prime}+i+F+1}
                              \left\{\begin{array}{ccc}
                                      j^{\prime} & F^{\prime} & i \\
                                      F          & j          & 1
                              \end{array}
                              \right\}.
\end{eqnarray}
Somewhat less general and elegant identities \cite{vmkh} relate
$$ \sqrt{i(i+1)(2i+1)} \left\{\begin{array}{ccc}
                                      j^{\prime} &  j   & 2 \\
                                      i          &  i   & 1 \\
                                      F^{\prime} &  F   & 1
                              \end{array}
                              \right\}$$
to
$$ \left\{\begin{array}{ccc}
                                      j^{\prime} & F^{\prime} & i \\
                                      F          & j          & 1
                              \end{array}
                              \right\}.$$

Finally, we come to the following results for NSD optical activity in bismuth.
\newline
Infrared line (876 nm):
\begin{eqnarray}\label{876}
\frac{P_a}{\kappa_a}=\frac{P_Q}{Q}\;
\frac{2\gamma + 1}{3}\;\frac{K}{i(i+1)}\;(a_1 a^{\prime}_2 +a_2 a^{\prime}_3 -
a_2 a^{\prime}_1
- a_3 a^{\prime}_2)^{-1}\nonumber \\
\cdot\Biggl\{(a_1 a^{\prime}_2 +a_2 a^{\prime}_3 + a_2 a^{\prime}_1 + a_3
a^{\prime}_2
- 2\sqrt{2/5}\, a_2 a^{\prime}_2)
                   \left(\begin{array}{c}
                                     F+1\\
                                      0 \\
                                      -F
                              \end{array}
                                 \right)\nonumber \\
+ (a_1 a^{\prime}_2 - a_2 a^{\prime}_3 - a_2 a^{\prime}_1
+ a_3 a^{\prime}_2) \frac{1}{2} \left(\begin{array}{c}
                                     (F+1)^2 - 26\\
       \frac{F^2(F+1)^2 - 52 F(F+1) + 546}{F(F+1) - 21} \\
                                      F^2 - 26
                              \end{array}
                                 \right)\Biggr\}\nonumber \\
= - 10^{-9}\Biggl\{0.645 \left(\begin{array}{c}
                                     F+1\\
                                      0 \\
                                      -F
                              \end{array}
                                 \right)
+ 0.274 \left(\begin{array}{c}
                                     (F+1)^2 - 26\\
       \frac{F^2(F+1)^2 - 52 F(F+1) + 546}{F(F+1) - 21} \\
                                      F^2 - 26
                              \end{array}
                                 \right)\Biggr\}.
\end{eqnarray}
Red line (648 nm):
\begin{eqnarray}\label{648}
\frac{P_a}{\kappa_a}=\frac{P_Q}{Q}\;
\frac{2\gamma + 1}{3}\;\frac{K}{i(i+1)}\;(- a_1 + a_3)^{-1}\nonumber \\
\cdot\Biggl\{(- a_1 - a_3 - 8\sqrt{2/5}\, a_2)
                   \left(\begin{array}{c}
                                     F - 3/2\\
                                       - 5/2\\
                                   - F - 5/2
                              \end{array}
                                 \right)
+ (- a_1 - a_3) \left(\begin{array}{c}
                                 - 2F^2 + F + 105/2\\
                                  - 2F^2 - 2F + 99/2 \\
                                  - 2F^2 - 5F + 99/2
                              \end{array}
                                 \right)\Biggr\}\nonumber \\
= 10^{-9}\Biggl\{0.370 \left(\begin{array}{c}
                                     F - 3/2\\
                                       - 5/2\\
                                   - F - 5/2
                              \end{array}
                                 \right)
- 0.780 \left(\begin{array}{c}
                                 - 2F^2 + F + 105/2\\
                                  - 2F^2 - 2F + 99/2 \\
                                  - 2F^2 - 5F + 99/2
                              \end{array}
                                 \right)\Biggr\}.
\end{eqnarray}
We use here for $P_Q$ the results of the phenomenological calculations
\cite{nsfk1,khr}
\begin{equation}\label{phbi1}
P_Q(876) = 2.9\cdot 10^{-7}(-Q/N),
\end{equation}
\begin{equation}\label{phbi2}
P_Q(648) = 3.8\cdot 10^{-7}(-Q/N)
\end{equation}
for the infrared and red line. In formulae (\ref{876}) and
(\ref{648}) the first, second and third lines in each column refer to the
transitions $F\longrightarrow F+1, F\longrightarrow F$ and $F\longrightarrow
F-1$ correspondingly.

The numerical predictions for all hyperfine components of the two transitions
are collected in Table 2. They are obviously in a reasonable agreement with
those of Ref.\cite{nsfk} which are included for comparison into the Table.
  \begin{table}[h]
  \begin{center}
  \begin{tabular}{cccccccccc}
   &            & &  &876 nm  &876 nm      &  & &648 nm  &648 nm      \\
   &            & &  &this    &\cite{nsfk}&  & & this   &\cite{nsfk}\\
   &            & &  &work    &        &    & & work   &          \\
   &            & &  &      &          &  & &        &         \\
$F$&$F^{\prime}$& &  &      &          &  & &        &        \\
6  & 7          & &  &--     & --      &   & & 1.22   &  1.01   \\
6  & 6          & &  &$-$0.17&$-$0.15   &   & & 2.60   &   2.39  \\
6  & 5          & &  & 0.11  &  0.08   &   & & 3.78   &   3.57  \\
5  & 6          &&&$-$0.66&$-$0.56    &  & &$-$0.46  &$-$0.48  \\
5  & 5          & &  &0.35  &  0.29    &  & &  0.73  &   0.70  \\
5  & 4          & &  &0.35  &  0.29    &  & &  1.71  &   1.69  \\
4  & 5          &&&$-$0.30&$-$0.24    &&&$-$1.82 &$-$1.69  \\
4  & 4          & & &$-$2.58  &$-$2.17    &  & &$-$0.83  &$-$0.70  \\
4  & 3          & & & 0.54  &  0.45    &  & & $-$0.05  &   0.09  \\
3  & 4          & & & 0.02  &  0.03    &  & &$-$2.87  &$-$2.61  \\
3  & 3          & & & 0.20  &  0.18    &  & &$-$2.08  &$-$1.82  \\
3  & 2          & & &--     &   --     &  & &$-$1.49  &$-$1.23  \\
        \end{tabular}
  \end{center}
  \caption{$(P_a/\kappa_a)\cdot 10^8$ at hyperfine components of infrared and
           red transitions}
  \label{su}
  \end{table}

\section{Accuracy of results}
It is proper now to compare the accuracy of the present results with that of
the results of Ref.\cite{nsfk} and with that of theoretical predictions for the
NSI optical rotation.

Let us start with thallium. Here in the phenomenological approach adopted in
Refs.\cite{nsfk1,nsfk} and here, both NSI and NSD weak interactions admix to
the initial state $6p_{1/2}$ exactly same simple excitations $ns_{1/2}$. As to
the excitations of the type $6s6p^2$, when treating both effects the energy
splitting between them was neglected. There is no special reasons to expect
that such an averaging can influence NSD and NSI effects in essentially
different ways. Therefore, for the accuracy of our thallium result (\ref{th})
one can accept the same estimate 15\% as was given in Refs.\cite{nsfk1,khr} for
the NSI prediction (\ref{phth}). It should be mentioned that, taking account of
the mentioned 15\% error, the phenomenological NSI result (\ref{phth}) agrees
well with the most accurate subsequent theoretical calculation \cite{dfs} which
gives for this transition $P_Q = 3.20(10)\cdot 10^{-7}(-Q/N)$. The experimental
result \cite{wbs}, $P_Q = 2.50(38)\cdot 10^{-7}$, does not contradict the
theoretical predictions (at $-Q/N=0.947$).

A close situation takes place in lead. Usual excitations of the $6p$ electron
belong to the configurations $ns\;6p$ which are well described by the $jj$
coupling approximation.  The NSI and NSD weak interactions admix to the
dominating, $|1/2\;1/2>_0$, component of the ground state wave function
states of different total electronic angular momentum $J$, $|\cdot\;1/2>_0$ and
$|\cdot\;1/2>_1$ respectively (here the dot denotes the $ns$-electron).
However, the energy interval between the states $|\cdot\;1/2>_0$ and
$|\cdot\;1/2>_1$ is tiny, about $300\; cm^{-1}$ even at $n=7$. In the
contribution to the effect of the second ground state component,
$|3/2\;3/2>_0$, the usual $ns$ excitations are admixed to the final state
$|^3 P_1>= |1/2\;3/2>_1$; the NSI weak interaction admixes the excitations
$|\cdot\;3/2>_1$, and the NSD one admixes both $|\cdot\;3/2>_1$ and
$|\cdot\;3/2>_2$. The energy interval between the last two states is also small
($1250\; cm^{-1}$ for $n=7$).  And finally as concerns the contributions of the
$6s$-electron excitations, the same arguments apply as in the case of Tl.
Therefore the averaged treatment of all the excited states when evaluating the
NSD effects in lead in Ref.\cite{nsfk} is practically of the same accuracy as
the phenomenological calculation of NSI effects in Ref.\cite{nsfk1}.

However, in the present approach an extra approximation is made as compared to
Ref.\cite{nsfk}. From the above argument concerning the admixture of $ns\;6p$
excitations it follows that to treat both ground state components on the same
footing one has to neglect the energy interval between the ground state and
$|^3 P_1>= |1/2\;3/2>_1$. But this interval, $7819\; cm^{-1}$, is not as
negligible. Fortunately, the ground state of lead is also close to pure $jj$
coupling, the relative admixture of the $|3/2\;3/2>_0$ component in it is small
(see (\ref{intc})). Thus the error introduced by that extra approximation gets
negligible, and the present result coincides numerically with that of
Ref.\cite{nsfk}. The accuracy of both can be estimated to be the same, 20\%, as
that of the phenomenological NSI one (\ref{phle}). It is worth noting here that
the phenomenological NSI result (\ref{phle}) agrees within the indicated error
with subsequent theoretical calculations \cite{bb,df} and experiment
\cite{mvm}.

In bismuth the situation is rather different. In spite of the reasonable
agreement between the present results and previous ones \cite{nsfk} (see Table
2), the neglect of the energy intervals between the states belonging to the
$6p^3$ configuration is more essential than in lead. Furthermore, in bismuth,
as distinct from thallium and lead, the chain
$$p_{1/2}\longrightarrow s_{1/2}\longrightarrow p_{1/2},$$
becomes operative. Therefore, not only $ns\longrightarrow 6p_{3/2}$ E1
transitions, but $ns\longrightarrow 6p_{1/2}$ ones contribute to the effect. In
our treatment we have tacitly assumed that the corresponding radial integrals
are equal. However, they are not, at least for the E1 transition
$7s\longrightarrow 6p$ \cite{nsfk1}.

The latter correction can be introduced easily into our consideration.
Numerical calculations \cite{sf} give the following values for the radial
integrals of $7s\longrightarrow 6p$ transitions (in the units of the Bohr
radius)
$$\rho_{3/2} = 2.2;\;\;\;\rho_{1/2} = 1.5$$
The relative contributions of the $7s\,6p^2$ excitations to the effect
constitute, according to Ref.\cite{nsfk1}, 30\% and 24\% for the infrared and
red transitions respectively. So, the effect discussed can be accounted for by
the introduction of the correction factors 0.89 into the term
$2\sqrt{2/5}\, a_2 a^{\prime}_2$ in formula (\ref{876}) and 0.92 into the
term $8\sqrt{2/5}\, a_2$ in formula (\ref{648}). Both those terms are due
to the $ns\longrightarrow 6p_{1/2}$ E1 transitions.

The present results for Bi were recalculated at the standard set of the
intermediate-coupling coefficients $a_i$ (see Table 1) used in Ref.\cite{nsfk},
and the above correction was made, accounting for the difference between
$\rho_{3/2}$ and $\rho_{1/2}$. The numbers obtained in this way are indeed even
more close to the results of Ref.\cite{nsfk}.

A specific consequence of the neglect of the energy splitting inside the
$6p^3$ configuration is the absence in expression (\ref{ch1}) of the
contribution $$i_{+}\;\frac{b^{\dagger}_{+}b_{+} +
b^{\dagger}_{-}b_{-}}{\sqrt{2}}$$ scalar in electron variables. Such terms
arise at $q$ both in $\hat{D}^{+}\hat{W}$ and $\hat{W}\hat{D}^{+}$, but cancel
out in their sum. The cancellation is no more exact if the energy splitting
does not vanish. The comparison with the results of Ref.\cite{nsfk} demonstrate
that the scalar contribution to the infrared transition does not exceed few
percent. Obviously, it cannot contribute to the red transition with $\Delta J =
1$.

Of course, the aim and outcome of the above procedures is to establish the
correspondence between the two results for bismuth, to have one more check for
both of them. Clearly, in bismuth the present calculation by itself is less
accurate than that of Ref.\cite{nsfk}.

But what is the accuracy of the latter? In spite of an extra approximation,
that of averaging over the positions of all intermediate states belonging to
the same configuration, its error should be about the same as that of the
phenomenological calculation \cite{nsfk1} of NSI optical activity in bismuth.
The accuracy of the latter was stated to be 20\%.  Indeed the phenomenological
result (\ref{phbi1}) for the infrared transition within this error contradicts
neither most refined theoretical investigations \cite{df,df1}, nor experimental
results \cite{hol,mac,mzw}.

As to the red transition, the situation is less clear. There is no agreement
between the phenomenological result (\ref{phbi2}) for it and that of the
Hartree-Fock calculation \cite{df1}: $P_Q = 1.5(1.0)\cdot 10^{-7}(-Q/N)$.
Moreover, there is a discrepancy between experiments at this line. One of
them \cite{bz} gives $P_Q = 4.04(0.54)\cdot 10^{-7}$, two other \cite{bir,war}
$1.56(0.36)\cdot 10^{-7}$ and $1.96(0.18)\cdot 10^{-7}$ respectively. Both
discrepancies, theoretical and experimental, still are not resolved.

\bigskip
I am extremely grateful to S.K. Lamoreaux and P.G. Silvestrov for useful
discussions. This investigation was financially supported by the Russian
Fundamental Research Foundation, Grant No. 94-02-03942.

\pagebreak


\begin{thebibliography}{99}

\bibitem{fk} V.V. Flambaum, I.B. Khriplovich, Zh.Eksp.Teor.Fiz. {\bf 79}
(1980) 1656 [Sov.Phys. JETP {\bf 52} (1980) 835]
\bibitem{fks} V.V. Flambaum, I.B. Khriplovich, O.P. Sushkov, Phys.Lett.
{\bf B145} (1984) 367
\bibitem{nsfk} V.N. Novikov, O.P. Sushkov, V.V. Flambaum, I.B. Khriplovich,
Zh.Eksp.Teor.Fiz. {\bf 73} (1977) 802 [Sov.Phys. JETP {\bf 46} (1977) 420]
\bibitem{wie} M.S. Noecker, B.P. Masterson, C.E. Wieman, Phys.Rev.Lett.
{\bf 61} (1988) 310
\bibitem{mzw} M.J.D. Macpherson, K.R. Zetie, R.B. Warrington, D.N. Stacey,
J.P. Hoare, Phys.Rev.Lett. {\bf 67} (1991) 2784
\bibitem{mvm} D.M. Meekhof, P. Vetter, P.K. Majumder, S.K. Lamoreaux,
E.N. Fortson, Phys.Rev.Lett. {\bf 71} (1993) 3442
\bibitem{lam} S.K. Lamoreaux, Talk at the XIV Moriond Workshop, January 1994
\bibitem{frk} P.A. Frantsuzov, I.B. Khriplovich, Z.Phys. {\bf D7} (1988) 297
\bibitem{kra} A.Ya. Kraftmakher, Phys.Lett. {\bf A132} (1988) 167
\bibitem{bjs} S.A. Blundell, W.R. Johnson, J. Sapirstein, Phys.Rev.Lett.
{\bf 65} (1990) 1411
\bibitem{bp} C. Bouchiat, C.A. Piketty, Z.Phys. {\bf C49} (1991) 91
\bibitem{khr} I.B. Khriplovich, Parity Nonconservation in Atomic Phenomena
(Gordon and Breach, London, 1991)
\bibitem{dkt} V.F. Dmitriev, I.B. Khriplovich and V.B. Telitsin,
Nucl.Phys.A, in press; preprint BINP 93-115
\bibitem{fk1} V.V. Flambaum, I.B. Khriplovich, Zh.Eksp.Teor.Fiz. {\bf 89}
(1985) 1505 [Sov.Phys. JETP {\bf 62} (1985) 872]
\bibitem{koz} M.G. Kozlov, Phys.Lett. {\bf A130} (1988) 426
\bibitem{bp1} C. Bouchiat, C.A. Piketty, Phys.Lett. {\bf B269} (1991) 195;
erratum {\bf B274} (1992) 526
\bibitem{nsfk1} V.N. Novikov, O.P. Sushkov, V.V. Flambaum, I.B. Khriplovich,
Zh.Eksp.Teor.Fiz. {\bf 71} (1976) 1665 [Sov.Phys. JETP {\bf 44} (1976) 872]
\bibitem{sfk1} O.P. Sushkov, V.V. Flambaum, I.B. Khriplovich,
Talk at the Conference on the Theory of Atoms and Molecules (Vilnius, 1979)
\bibitem{ll} D.A. Landman, A. Lurio, Phys.Rev. {\bf A1} (1970) 1330
\bibitem{vmkh} D.A. Varshalovich, A.N. Moskalev, V.K. Khersonsky, Quantum
Theory of Angular Momentum, pp. 306,321,322 (Nauka, Leningrad, 1975)
\bibitem{dfs} V.A. Dzuba, V.V. Flambaum, P.G. Silvestrov, O.P. Sushkov,
J.Phys. {\bf B20} (1987) 3297
\bibitem{wbs} T.D. Wolfenden, P.E.G. Baird, P.G.H. Sandars, Europhys.Lett.
{\bf 15} (1991) 731
\bibitem{bb} C.P. Botham, S.A. Blundell, A.-M. M{\aa}rtensson-Pendrill,
P.G.H. Sandars, Phys.Scr. {\bf 36} (1987) 481
\bibitem{df} V.A. Dzuba, V.V. Flambaum, P.G. Silvestrov, O.P. Sushkov,
Europhys.Lett. {\bf 7} (1988) 413
\bibitem{sf} V.V. Flambaum, O.P. Sushkov, J.Quant.Spectr.Rad.Transf. {\bf 20}
(1978) 569
\bibitem{df1} V.A. Dzuba, V.V. Flambaum, P.G. Silvestrov, O.P. Sushkov,
Phys.Lett. {\bf A141} (1989) 147
\bibitem{hol} J.H. Hollister {\it et al}, Phys.Rev.Lett. {\bf 46} (1981) 643
\bibitem{mac} M.J.D. Macpherson {\it et al}, Europhys.Lett. {\bf 4} (1987) 811
\bibitem{bz} L.M. Barkov, M.S. Zolotorev, Zh.Eksp.Teor.Fiz. {\bf 79}
(1980) 713 [Sov.Phys. JETP {\bf 52} (1980) 360]
\bibitem{bir} G.N. Birich {\it et al}, Zh.Eksp.Teor.Fiz. {\bf 87}
(1984) 776 [Sov.Phys. JETP {\bf 60} (1984) 442]
\bibitem{war} R.B. Warrington, C.D. Thomson, D.N. Stacey, Europhys.Lett.
{\bf 24} (1993) 641

\end{thebibliography}
\end{document}